\documentclass[structabstract]{aa}
\usepackage{natbib}                     % BibTeX style bibliography
\usepackage{amsmath}                    % Mathematic formulaes
\usepackage{graphicx}                   % Pictures and images
\usepackage[varg]{txfonts}              % A&A is printed using the Postscript TX Times-fonts.
\usepackage[english]{babel}             % Language is english
\usepackage{nameref}                    % References
\usepackage[normalem]{ulem}             % for striketrough type effects
\usepackage{paralist}                   % lists
\usepackage{multirow}
\usepackage{color}

\begin{document}

  \title{Constraining planet structure and composition from stellar chemistry: trends in different stellar populations}

  \author{N. C. Santos\inst{1,2}
%       \and  several nice people 
%       \and from here and abroad
          \and V. Adibekyan\inst{1}
          \and C. Dorn\inst{3}
          \and C. Mordasini\inst{4}
          \and L. Noack\inst{5,6}
          \and S. C. C. Barros\inst{1}
          \and E. Delgado-Mena\inst{1}
          \and O. Demangeon\inst{1}
          \and J. Faria\inst{1,2}
%         \and P. Figueira\inst{1}
          \and G. Israelian\inst{7,8}
          \and S. G. Sousa\inst{1}
         }

  \institute{
          Instituto de Astrof\'isica e Ci\^encias do Espa\c{c}o, Universidade do Porto, CAUP, Rua das Estrelas, 4150-762 Porto, Portugal
          \and
          Departamento de F\'isica e Astronomia, Faculdade de Ci\^encias, Universidade do Porto, Rua do Campo Alegre, 4169-007 Porto, Portugal
         \and
         %Center for space and habitability, Physikalisches Institut, University of Bern, Sidlerstrasse 5, 3012, Bern, Switzerland
         University of Zurich, Institut of computational sciences, Winterthurerstrasse 190, CH -8057 Z\"urich
         \and
         Physikalisches Institut, University of Bern, Gesellschaftsstrasse 6, CH-3012 Bern, Switzerland 
         \and
         Department of Reference Systems and Geodynamics, Royal Observatory of Belgium (ROB), Avenue Circulaire 3,1180 Brussels, Belgium
         \and
         Institute of Geological Sciences, Freie Universit\"at Berlin, Malteserstr. 74-100, 12249 Berlin, Germany 
         \and
         Instituto de Astrof\'isica de Canarias, C/V\'ia L\'actea s/n, 38205 La Laguna, Tenerife, Spain
         \and 
         Universidad de La Laguna, Dept. Astrof\'isica, E-38206 La Laguna, Tenerife, Spain
}

  \date{Received date / Accepted date }
%----------------------------------------------------------------------------------------
%       Abstract
%----------------------------------------------------------------------------------------
  \abstract
  {The chemical composition of stars that have orbiting planets provides important clues about the frequency, architecture, and composition of exoplanet systems. }
  {We explore the possibility that stars from different galactic populations that have different intrinsic abundance ratios may produce planets with a different overall composition.}
  {We compiled abundances for Fe, O, C, Mg, and Si in a large sample of solar neighbourhood stars that belong to different galactic populations. We then used a simple stoichiometric model to predict the expected iron-to-silicate mass fraction and water mass fraction of the planet building blocks, as well as the summed mass percentage of all heavy elements in the disc.}
  {Assuming that overall the chemical composition of the planet building blocks will be reflected in the composition of the formed planets, we show that according to our model, {discs around} stars from different galactic populations, as well as {around} stars from different regions in the Galaxy, are expected to form rocky planets with significantly different iron-to-silicate mass fractions. The available water mass fraction also changes significantly from one galactic population to another. }
  {The results may be used to set constraints for models of planet formation and chemical composition. Furthermore, the results may have impact on our understanding of the frequency of planets in the Galaxy, as well as on the existence of conditions for habitability.}

%----------------------------------------------------------------------------------
%       Title
%----------------------------------------------------------------------------------

  \maketitle
%---------------------------------
  
  -------------------------------------------------
%       Introduction
%----------------------------------------------------------------------------------
  \section{Introduction}                                        \label{sec:Introduction}

% Caroline Dorn:
% I would stress in the letter that the strength of the model is to simplify the condensation sequence for building blocks from which planets are formed. 
% Then in the discussion, we can discuss what the knowledge of condensed material tell us about the make-up of planets and maybe interior dynamics. 
% This is to acknowledge planet differentiation and core-formation processes that may alter the actual mineralogical compounds of a planet.

The study of stars hosting planets is providing a huge amount of information about the processes of planetary formation and evolution \citep[see e.g.][]{Mayor-2014}. The dependence on the frequency of planets with the stellar metallicity and mass \citep[e.g.][]{Santos-2004b,Fischer-2005,Johnson-2007,Sousa-2011,Buchhave-2012}, for example, has been suggested as strong evidence in favor of
the hypothesis
of the core-accretion model as the dominant giant planet formation process \citep[e.g.][]{Mordasini-2012}. The architecture of planetary systems and its dependence on the metal content of the stars further provides indications about the processes involved in the planet migration \citep[e.g.][]{Dawson-2013, Adibekyan-2013,Beauge-2013}.

The stars we observe in the solar neighbourhood can be divided into three galactic populations: the thin-disc, the thick-disc, and the halo population. Most stars are members of the younger thin-disc component, ranging in [Fe/H] from $\sim$$-0.8$ up to $\sim$+0.5\,dex. Thick-disc stars typically have lower metallicities than their thin-disc counterparts. More specifically, they present higher values of $\alpha$-element\footnote{Elements for which the most abundant isotopes are integer multiples of 4, the mass of a helium nucleus ($\alpha$ particle)} abundances \citep[for a recent paper, see][]{Adibekyan-2013b}. Finally, halo stars are usually objects of lower metallicity, which also often present $\alpha$ element enhancement. They are commonly identified using dynamical approaches \citep[][]{Bensby-2003}. The kinematical and chemical properties (in particular the abundance ratios) of these three populations reflect their origin, age, and the galactic formation process \citep[e.g.][]{Haywood-2013}; see Appendix\,\ref{appendix1A} for more details.

Recent studies suggest that the abundances of specific chemical species in the stellar photosphere may give clues about the internal structure and composition of the planets. This is true both for giant planets \citep[][]{Guillot-2006,Fortney-2007} and for their rocky counterparts \citep[e.g.][]{Bond-2010,Delgado-Mena-2010,Dorn-2015,Thiabaud-2015,Santos-2015a,Dorn-2017}. Stars from different galactic populations may thus {present} different planet {frequencies, and the planets orbiting them may present different composition trends} \citep[e.g.][]{Haywood-2008,Adibekyan-2012,Frank-2014,Adibekyan-2015b,Adibekyan-2016}. 

In this context, the present paper investigates whether stars that come from different galactic populations and have different chemical composition are expected to form {planet building blocks (or planets)} with different compositions. In Sect.\,\ref{sec:data} we present the data selected for this study, including detailed chemical abundances for several elements. In Sect.\,\ref{sec:model} we then describe our model, applying it to the stars in different galactic populations in Sect.\,\ref{sec:composition}. Finally, in Sect.\,\ref{sec:conclusions} we discuss our results in face of the expected planet populations in the galaxy, including the prospects for life-bearing worlds.

  \section{Data}                                \label{sec:data}

To explore the effect of different initial stellar abundances on the planet composition, we need to define a sample of stars for which precise abundances have been determined. As we show below, our model needs abundance values for Fe, Si, Mg, C, and O as input. To build this sample, we started from the work of \citet[][]{Adibekyan-2012b}. This study, based on spectra with
high signal-to-noise (S/N)
ratios and high resolution obtained with the HARPS spectrograph, provides abundances of Fe, Si, and Mg for 1111 stars in a volume-limited sample. Oxygen abundances were then added from the study of \citet[][]{BertrandeLis-2015}. The values derived using the 6158\,\AA\ oxygen line were preferred, as these have been shown  {by the authors} to be more precise. Finally, for C we used the carbon values recently derived by \citet[][]{Suarez-Andres-2017}. All these abundances were derived based on same set of uniform stellar atmospheric parameters \citep[namely T$_{eff}$ and $\log{g}$, from][]{Sousa-2008,Sousa-2011,Sousa-2011b}.
%Finally, for C we used the relations described in \citep[][their equation 4]{Nissen-2014} to estimate the [C/H] values as a function of stellar metallicity.

All abundances listed in these papers (as well as the stellar parameter analysis) were computed relative to the Sun. They were transformed into absolute abundances assuming the solar composition as given in \citet{Asplund-2009} for Fe ($\log{\epsilon}$=7.50), Mg ($\log{\epsilon}$=7.60), and Si ($\log{\epsilon}$=7.51), and as given by \citet[][]{BertrandeLis-2015} and \citet[][]{Suarez-Andres-2017} for O ($\log{\epsilon}$=8.71) and C ($\log{\epsilon}$=8.50).\footnote{The value [X/H], where X is a specific element, is defined as [X/H] = $\log{\frac{N_X}{N_H}}$ - $\log{(\frac{N_X}{N_H})}_\odot$, where $N$ is the number of atoms. The values of $\log{\epsilon}$ are defined as $\log{\epsilon}$=$\log{\frac{N_X}{N_H}}+12$.} {For helium, the adopted value was taken from \citet[][$\log{\epsilon}$=10.93]{Lodders-2003}.}

In total, we have 535 stars for which precise values of abundances for Fe, Si, Mg, C, and O have been obtained. These stars represent the solar neighbourhood sample well. In order to avoid strong systematic effects or errors in the abundances caused by uncertainties in the line-lists and model atmospheres as we move away from the solar values \citep[see e.g.][]{Adibekyan-2012b,BertrandeLis-2015}, we decided to further cut our sample to include only stars with a temperature of $\pm$300\,K around solar (we assume T$_{eff}(\odot)$=5777\,K). This left us with 371 stars.

To classify the stars into different galactic populations, we used the chemical boundary discussed in \citet[][]{Adibekyan-2012b}. According to this, 303 of the 371 stars belong to the thin disc, while 68 have enhanced $\alpha$-element abundances. This latter pattern is typical of thick-disc stars of our Galaxy. However, for $\alpha$-rich metal-rich stars ([Fe/H]>$-0.2\,dex$), there is no consensus about which galactic population they belong to \citep[][]{Adibekyan-2011,Bensby-2014}. We therefore divided the 68 stars into two different groups: the thick-disc population (48 stars with [Fe/H]<$-0.2$\,dex) and its $\alpha$-rich metal-rich counterpart (20 stars with [Fe/H]$\geq-0.2$\,dex), hereafter called h$\alpha$mr. Finally, 3 stars in our sample were identified as belonging to the galactic halo following the kinematic criteria of \citet[][]{Bensby-2003}. They were treated separately since halo stars in the solar neighbourhood are also known to be mainly rich in $\alpha$ elements \citep[but see][]{Nissen-2010}.

  \section{Model}                           \label{sec:model}

The model we used here is the same as was used in \citet[][]{Santos-2015a}. In brief, it makes use of the abundances of the rock-forming elements Fe, Si, Mg, C, and O, together with H and He, and assumes that these are the most relevant to control the species expected from equilibrium condensation models \citep{Lodders-2003,Seager-2007}, such as H$_{2}$, He, H$_{2}$O, CH$_4$, Fe, MgSiO$_{3}$, Mg$_{2}$SiO$_{4}$, and SiO$_2$\footnote{There are more silicate forms, but these are the most relevant.}. {In other words, in our model 
we only include the mineral phases of the main rock-forming elements that dominate the crust, the upper and lower mantle, and the
core of an Earth-like planet interior \citep[see e.g.][]{McDounough-1995,Sotin-2007}}. A simplified model for the expected mass fractions of different compounds using these species is thus a reasonable approach. In this case, the molecular abundances and therefore the mass fraction can be found from the atomic abundances with simple stoichiometry, as discussed in \citet[][]{Santos-2015a}; see also \citet[][]{Bond-2010,Thiabaud-2015,Unterborn-2016}.

We note that no star in our sample has values of Mg/Si>2, in which case, Si would be incorporated in olivine and the remaining Mg would enter in other minerals, mostly oxides. Our simple model does not take these cases into consideration. Moreover, ten stars were found to have C/O ratios above 0.8: above this value, the mineralogy is expected to be significantly different \citep[][]{Bond-2010}, with carbides forming instead of the silicates; planet building blocks would then be strongly enriched in carbon. Since only one star was found with C/O above 1 (C/O=1.14), and given the typical (high) errors in the derivation of abundances for these species \citep[see][]{BertrandeLis-2015,Suarez-Andres-2017}, we decided to keep these stars in the sample. In any case, observations suggest that only a small fraction of the stars in the solar neighbourhood have C/O>0.8 \citep[][]{Fortney-2012,Brewer-2016}.
%\vspace{-0.1truecm}

The simple relations presented in \citet[][]{Santos-2015a} {(see also Appendix\,\ref{appendix1B})} allow us to compute the expected mass fractions, in particular, the iron-to-silicate mass fraction ($f_{\rm iron}$)\footnote{Not taking into account the iron silicates.}, the water mass fraction ($w_f$), and the summed mass percent of all heavy elements ($Z$) expected for the planetary building blocks/grains:
\vspace{-0.15truecm}
\begin{eqnarray}
f_{\rm iron}=m_{\rm Fe}/(m_{\rm Fe}+m_{\rm MgSiO3}+m_{\rm Mg2SiO4}+m_{\rm SiO2})\\
w_f=m_{\rm H2O}/(m_{\rm H2O} +m_{\rm Fe}+m_{\rm MgSiO3}+m_{\rm Mg2SiO4}+m_{\rm SiO2})\\
\begin{aligned}
Z=(m_{CH4}+m_{H2O}+m_{Fe}+m_{MgSiO3}+\\m_{Mg2SiO4}+m_{SiO2})/M_{tot}
\end{aligned}
,\end{eqnarray}
{where $m_X=N_x*\mu_X$,
%\begin{equation}
$M_{tot}=N_H*\mu_H+N_{He}*\mu_{He}+N_C*\mu_C+N_O*\mu_O+N_{Mg}*\mu_{Mg}+N_{Si}*\mu_{Si}+N_{Fe}*\mu_{Fe}$, and $N_X$ represents the number of atoms of each species $X$, and $\mu_X$ their mean molecular weights. All $N_X$ values are computed relative to hydrogen.}
%\end{equation}

These variables are expected to have an important influence on the planet structure and radius \citep[][]{Dorn-2017}. They are also expected to provide relevant information about some parameters relevant for the planet formation process, such as the mass fraction of all heavy elements and the ice mass fraction in solid planets, or giant planet cores that form beyond the water ice-line. 

We should note that in the calculations above we assume that the composition of H, C, N, O, Mg, Si, and Fe dominates the composition of the solid material. This is likely a good approximation, as in the Sun these elements represent the main contributors to the overall composition \citep[see abundances in e.g.][]{Asplund-2009}. 

The model predicts that planet building blocks for solar system have $f_{iron}=33\%$, $w_f=60\%$, and $Z=1.26\%$. The value of $f_{iron}$ is compatible to the value observed in the terrestrial planets (see the discussion below), while $w_f$ and $Z$ are similar to the values derived in \citet[][67.11\% and 1.31\%, respectively]{Lodders-2003}.

In Appendix\,\ref{appendix1C} we present some simple relations to compute the values of $f_{iron}$ and $Z$ based on the stellar abundances. 

%I would rather say that this might be a good assumptions since relative bulk composition of refractories are correlated between meteorites, and the solar photosphere, and maybe also the terrestrial planets of the Solar System, e.g. Earth and Mars. ...you could cite Sotin & Grasset 2007 for your previous sentence (e.g., "We note that..") But the Thiabaud reference is not a strong argument.

\section{Results} \label{sec:composition}

\begin{figure*}[t!]
\begin{center}
\begin{tabular}{c}
\includegraphics[angle=0,width=0.5\linewidth]{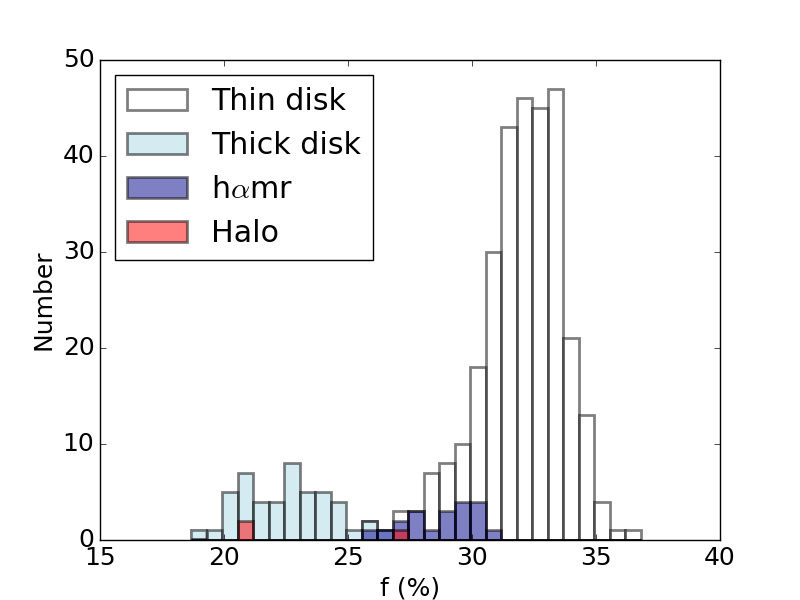}
\includegraphics[angle=0,width=0.5\linewidth]{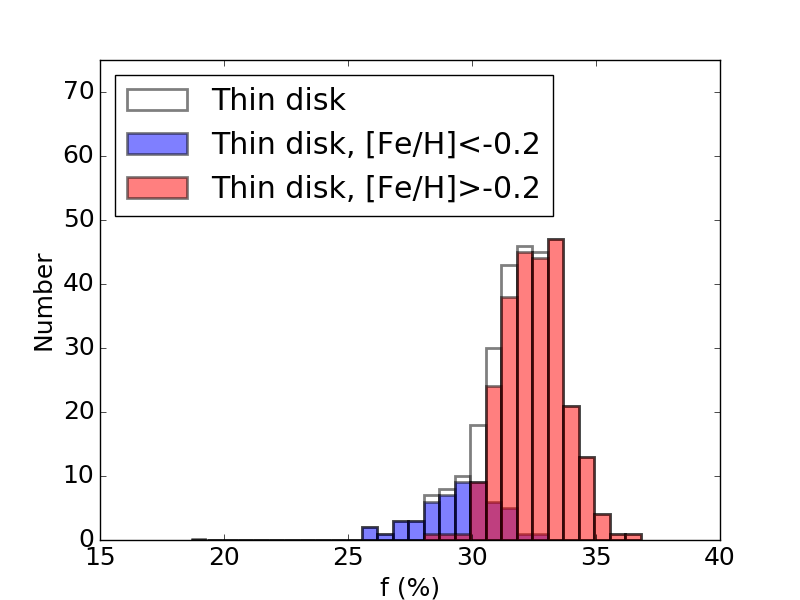}\\[-0.0truecm]
\includegraphics[angle=0,width=0.5\linewidth]{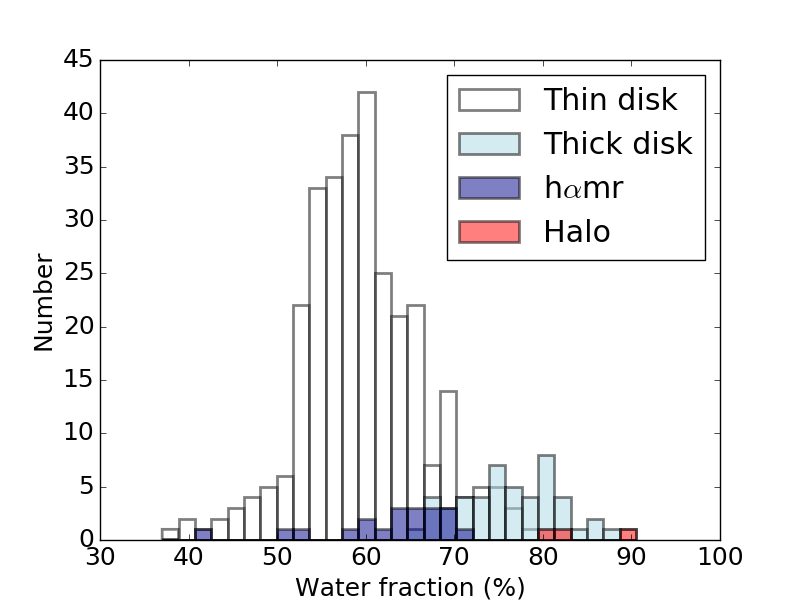}
\includegraphics[angle=0,width=0.5\linewidth]{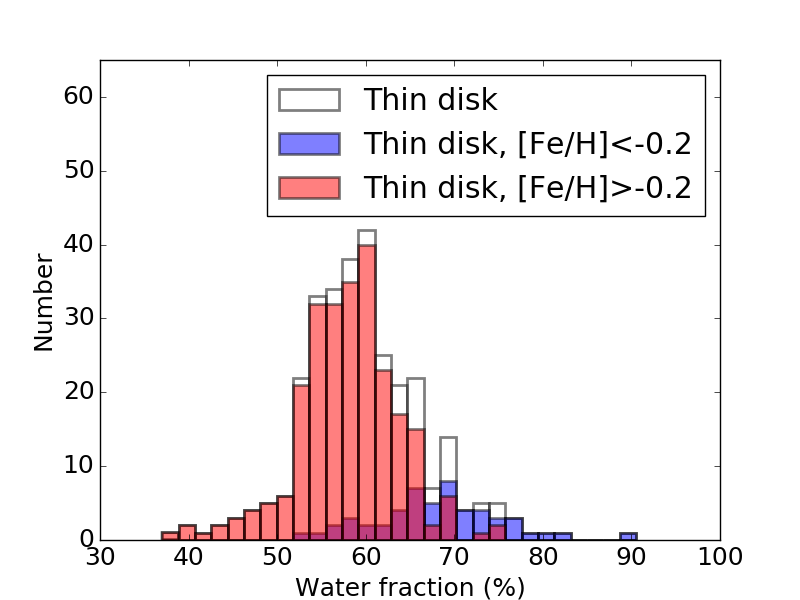}\\[-0.0truecm]
\includegraphics[angle=0,width=0.5\linewidth]{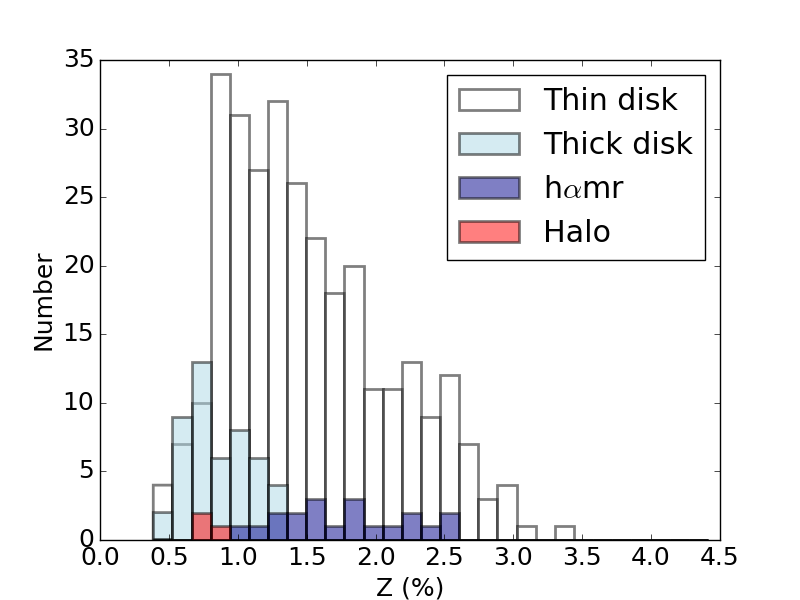}
\includegraphics[angle=0,width=0.5\linewidth]{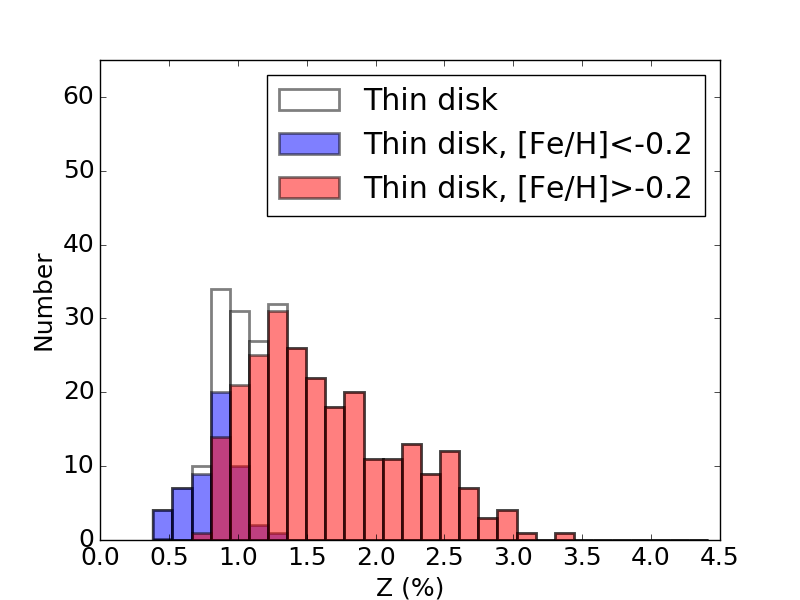}
\end{tabular}
\end{center}
\vspace{-0.5cm}
\caption{$f_{iron}$ (upper panels), $w_f$ (middle panels), and $Z$ (lower panels) mass fractions distributions for the different populations in the Galaxy.}
\label{fig:pop}
\end{figure*}

In Fig.\,\ref{fig:pop} we present the distributions of iron and water mass fractions (top and middle panels) for the stars in our sample, as derived with the model described above. The bottom panels present the values of $Z$. The left panels present the comparison of the four galactic populations mentioned above, and the right panels present the comparison between thin-disc stars of different metallicity. This separation is motivated by the fact that low-metallicity thin-disc stars present an increasing $\alpha$-element enhancement as [Fe/H] decreases \citep[e.g.][]{Adibekyan-2012}. In Table\,\ref{tab:pop} we present the average and standard deviation values for the different populations.

\begin{table}[t]
\caption{Average values and standard deviations for $f_{\rm iron}$ and the water mass fraction in the different galactic populations. }
\begin{center}
\begin{tabular}{lcc}
\hline
Population      & Average & Std.\\
\hline
$f_{\rm iron}$ & \\[0.1truecm]
Thin Disc & 31.974& 1.750\\
Thin Disc ([Fe/H]<$-$0.2) & 29.430 & 1.550\\
Thin Disc ([Fe/H]$\geq-$0.2) & 32.514 & 1.240\\
Thick Disc ([Fe/H]<$-$0.2)& 22.464 & 1.742\\
h$\alpha$mr&28.727 & 1.338\\
Halo$^\dagger$ & 23.110 & 2.884\\
\hline
$w_f$ &  \\[0.1truecm]
Thin Disc & 59.713 & 7.106\\
Thin Disc ([Fe/H]<$-$0.2) & 68.014& 7.116\\
Thin Disc ([Fe/H]$\geq-$0.2) & 57.953 & 5.724\\
Thick Disc ([Fe/H]<$-$0.2)&76.172& 5.448\\
h$\alpha$mr&62.591& 7.194\\
Halo$^\dagger$ &83.990 & 4.115\\
\hline
$Z$ &  \\[0.1truecm]
Thin Disc & 1.508 & 0.597\\
Thin Disc ([Fe/H]<$-$0.2) & 0.818 & 0.181\\
Thin Disc ([Fe/H]$\geq-$0.2) & 1.655 & 0.550\\
Thick Disc ([Fe/H]<$-$0.2)&0.870 & 0.233\\
h$\alpha$mr&1.773 & 0.444\\
Halo$^\dagger$ &0.808 & 0.082\\
\hline
\end{tabular}
\newline
$\dagger$ We only have three halo stars, which limits the conclusions for this specific galactic population.
\end{center}
\label{tab:pop}
\end{table}%

The results presented in Fig.\,\ref{fig:pop} and in the table show that (on average) we can expect planet building blocks in different galactic populations to have significantly different values of iron and water mass fraction. Average values of $f_{iron}$ over the different populations vary between 22.5 and 23.1 for the metal-poor thick-disc and halo stars, respectively, and rise up to 32.5 for the metal-rich thin-disc population. Water mass fractions also vary from average values of $\sim$58\% for the thin-disc metal-rich stars to 76\% and 83\% for the metal-poor thick-disc and halo. 

Several conclusions can be drawn from comparing different groups of thin-disc stars. First, and as expected, the value of $f_{iron}$ and $w_f$ for stars of the metal-rich thin-disc population (where the Sun is included) suggest that these should be able to {host} planets similar in composition to the solar system planets. Metal-poor thin-disc stars, on the other hand, are expected to have lower iron and higher water mass fractions than the fractions expected for the Sun. The relatively small increase in the water fraction is related to the relatively higher [O/Fe] ratios observed in metal-poor thin-disc stars \citep[][]{BertrandeLis-2015} when compared to their metal-rich counterparts. The lower $f_{iron}$ reflects the decrease in metallicity and increase in the $\alpha$-element abundances.

According to our results, the metal-poor thick-disc population is expected to produce planets with much lower iron mass fractions (that may be translated into smaller cores). The water mass fraction in these stars, however, is higher than the fraction found in thin-disc objects. This is again expected as these stars present on average higher oxygen abundances than thin-disc stars of the same [Fe/H]. Interestingly, our results also suggest that h$\alpha$mr stars might produce planets with values of $f_{\rm iron}$ and $w_f$ intermediate between the thin-disc and the metal-poor thick-disc stars. This reflects their high $\alpha$-element abundances (O, C, Si, and Mg) when compared to thin-disc stars of the same metallicity, which leads to a relative increase in water mass fraction but a decrease of $f_{iron}$.

We only have three halo stars in our sample, which limits our conclusions regarding this population. However, and as expected, our model indicates that they might produce planets with a low iron mass fraction while being water rich. To our knowledge, there is no planet detected so far that orbits a halo star.

Finally, the average values of $Z$ for the different populations range from $\sim$0.8\% to $\sim$1.7\% (a factor of $\sim$2), with the individual values ranging from $\sim$0.5\% up to $\sim$3\%. As expected, metal-poor thin-disc and halo stars have the lowest average values of $Z$, while the remaining populations have higher values, even if a large spread is observed in each population. $Z$ is expected to be related to the planet formation efficiency. Increasing $Z$ may increase the probability of forming terrestrial-like planets or the icy/rocky cores of giant planets. This fact is behind the usual interpretation for the clear metallicity-giant planet correlation that is observed \citep[e.g.][]{Santos-2004b,Fischer-2005}. Interestingly, even if such a clear correlation has not been found for stars hosting the lower mass planets \citep[e.g.][]{Sousa-2011,Buchhave-2012}, hints exists that the overall metallicity still plays a role in this planet mass regime \citep[e.g.][]{Adibekyan-2012,Wang-2015,Zhu-2016}.

\section{Discussion: planet composition, frequency, and habitability}
\label{sec:conclusions}

The results presented above suggest that solar neighbourhood stars from different galactic populations whose abundance ratios are different may be able to produce planets, or at least the planet building blocks, that have different intrinsic compositions. Thick-disc stars in our Galaxy, for example, are known to be older than their thin-disc counterparts \citep[e.g.][]{Haywood-2013}. Stars in different galactic regions (inner vs. outer disk and stars with different heights above the galactic plane) also present different abundance ratios \citep[e.g.][]{Hayden-2015,Haywood-2016}. As has been discussed in some recent works \citep[e.g.][]{Frank-2014,Adibekyan-2016,Adibekyan-2016b}, this may imply that planets formed in different galactic ages (past or future) or in different galactic regions may have a significantly different composition trend. The planet formation efficiency {around} different populations of stars may also be different \citep[][]{Adibekyan-2012}.

Understanding whether the initial conditions for planet formation depend on the galactic population can also constrain the modelling of planet formation and planet structure. For instance, the overall amount of solid material ($Z$) and the expected composition of the building blocks ($f_{iron}$ and $w_f$) can be used as a prior to understand the frequency and populations of planets existing in the solar neighbourhood from planet population synthesis models \citep[e.g.][]{Mordasini-2012b}. Furthermore, priors on the planet composition can be set when modelling a planet based on its mass-radius relation \citep[e.g.][]{Dorn-2017}, as long as the abundances for some elements are known in the host star, or if the stellar population is deduced from kinematic studies (e.g. with GAIA),
for example. 

We note that our basic assumption is that the chemical composition of the (rocky) planets is intimately related to the chemical composition of the star. This is likely a good approximation from a statistical approach, as in the solar system the relative amount of refractory elements is correlated between meteorites, the solar photosphere, and the terrestrial planets Earth, Venus, and Mars \citep[e.g.][]{Morgan-1980,Lodders-2003,Drake-2002,Lodders-1998,Khan-2008,Sanloup-1999}. This is also found from models \citep[see e.g.][]{Thiabaud-2015}, even if exceptions may exist \citep[][]{Dorn-2017}. Finally, observations of rocky exoplanets also seem to suggest that they tend to follow the Earth composition line \citep[e.g.][]{Motalebi-2015}, in agreement with the assumptions taken here \citep[see also][]{Santos-2015a}. 
In this context, it would now be interesting to understand, for instance, if the mentioned result is compatible with the radius of planets orbiting metal-poor thick-disc stars \citep[e.g. {EPIC 210894022b/K2-111b} --][]{Fridlund-2017}. Exceptions to this trend may also have been found. These can be the cases of LHS\,1140\,b \citep[][]{Dittmann-2017}, for which the best-fit model implies a higher core mass fraction (even if compatible with Earth-like fractions within the errors), or the recently discovered very short period K2-106\,b \citep[][]{Guenther-2017}. {On the positive side, recent results suggest that the star TRAPPIST-1 may be a thick-disc object \citep[][]{Burgasser-2017}. Curiously, the planets discovered in this system that have a good mass estimate have been proposed to have a low density \citep[favouring a volatile-rich composition --][]{Gillon-2017}, a result compatible with our model.} 
It would be very interesting to investigate the detailed abundances of these specific stars in face of the results obtained above. New planet detections around bright nearby stars, expected with missions such as TESS, CHEOPS, and PLATO, will certainly allow us to address this field.

{
We also assume here that no internal or external physical process are able to significantly alter the initial stellar composition. Planet engulfment processes have been suggested to be responsible for specific stellar abundance patterns \citep[][]{Israelian-2001}, even if such events may be rare. Furthermore, the magnitude of this effect on the stellar abundances of G-type stars (with deep convective envelopes) is expected to be small when compared with the typical uncertainties in the abundances \citep[e.g.][]{Santos-2003}. Other less well understood effects, such as element diffusion, could also affect specific stellar abundances \citep[e.g.][]{Onehag-2014}, but the magnitude of this effect is in any case expected to be small for solar-type stars \citep[][]{Onehag-2014,Gonzalez-2014}. These effects need to be kept in mind, however, even if they may only be able to introduce minor changes to the original chemical abundances.
}

We also stress that the results discussed here can be used to estimate the composition of the planet building blocks, but they might not be directly applicable to the final composition or internal structure of a final differentiated planet (or chemical distribution within the planet). The composition of a planet of course depends on the region of the proto-planetary disc where it formed as well as on its migration path \citep[e.g.][]{Mordasini-2012b,Carter-Bond-2012}. Physical processes, including evaporation and collisional stripping, may also be able to change the overall composition of a planet (including its core-to-mantle fraction), as likely happened to Mercury \citep[][]{Benz-1988,Marcus-2010a}. 
Moreover, the compositional evolution of a planet depends on the accretion and differentiation history for which the understanding of the compositional model and the cooling history of the primitive body are key \citep[][]{Rubie-2015,Fischer-2017}. %(e.g., Rubie et al. 2015, Fischer et al. 2016 or 2017).
%Differentiation processes in the accreting planet -- separating the core from the mantle -- may also strongly depend on planet mass and may not occur for massive super-Earths \citep[][]{Elkins-Tanton2008}. 

Regarding the water content, and given the spread in water fractions in our solar system's terrestrial planets, we should also add that the origin of the water in our own planet Earth is still strongly debated, and several external processes (e.g. comet infall or strong stellar activity) that occurred during and after planet formation may likely significantly change the amount of water on the surface of a planet \citep[][]{Marcus-2010b,Luger-2015}. The values of $w_f$ derived are therefore expected to be indicative of the amount of water available in the proto-planetary disc; although this probably correlates with the final water content of the planets, strong scatter is expected. Water accumulates in the outer part of the disc during the later condensation of volatile atoms compared to heavier elements such as silicates and iron, which make up the main building blocks of the inner part of the proto-planetary disc. Even though the outer planets Jupiter, Saturn, Uranus, and Neptune contain mostly hydrogen, helium, and methane above their rocky core, the compositions of several of their moons are dominated by water and ice. 
{The amount of available water can also be related to the planet (or core) formation efficiency beyond the ice line, as it will control the amount of condensates in this region.}

Overall, our results suggest that stars in different galactic populations may produce planets with different core mass fractions and water contents. 
It is thus interesting to discuss the implications of this result for astrobiology. It has been suggested that large core mass fractions may inhibit plate tectonics \citep[][]{Noack-2014}. Plate tectonics is believed to be important for the evolution of complex life since it helps to maintain the magnetic field, replenishes the surface with nutrient-rich soil, and stabilizes the long-term global carbon cycle. Furthermore, in the absence of
plate tectonics, large cores can suppress outgassing of greenhouse gases on stagnant-lid planets \citep[][]{Noack-2017}. 
Finally, water has also been discussed to be a key ingredient for promoting plate tectonics \citep[][]{Korenaga-2010} and plays a crucial role for the surface habitability of rocky planets as well as for their interior evolution. Water in the mantle significantly reduces both silicate melting temperatures and mantle viscosity, and also influences the chemistry in the mantle.
Different planet compositions may in this context alter the width of the habitable zones, since plate tectonics is related to the capacity to degass CO$_2$. This may have implications for the definition of the so-called Galactic habitable zone \citep[e.g.][]{Lineweaver-2004,Prantzos-2008,Carigi-2013,Gonzalez-2014,Spitoni-2017}.
Since the quantity of available water in a planet may be related to the relative water content existing in the planet building blocks, the values derived in this paper may also give us some indications about the prevalence of water worlds \citep[][]{Leger-2004,Simpson-2017}.

%{\bf It is interesting to add that if confirmed, our results would have also implications for our understanding of planet atmospheres. For instance, recently \citet[][]{Lopez-2017} has shown that Ultra-Short Period planets could only retain their atmospheres if they are formed with a high-metallicity water rich envelope. Our model indicated that very-metal rich planets should be rare though.}

%----------------------------------------------------------------------------------------
%       AcknowledgementsPapersAndBooks
%----------------------------------------------------------------------------------------
\begin{acknowledgements}

This work was supported by Funda\c{c}\~ao para a Ci\^encia e a Tecnologia (FCT, Portugal) through the research grant through national funds
and by FEDER through COMPETE2020 by grants UID/FIS/04434/2013 \& POCI-01-0145-FEDER-007672, PTDC/FIS-AST/1526/2014 \& POCI-01-0145-FEDER-016886 and PTDC/FIS-AST/7073/2014 \& POCI-01-0145-FEDER-016880. P.F., S.B., N.C.S. e S.G.S. acknowledge support from FCT through Investigador FCT contracts nr. IF/01037/2013CP1191/CT0001, IF/01312/2014/CP1215/CT0004, IF/00169/2012/CP0150/CT0002, and IF/00028/2014/CP1215/CT0002. V.A. and E.D.M. acknowledge support from FCT through Investigador FCT contracts nr. IF/00650/2015/CP1273/CT0001, IF/00849/2015/CP1273/CT0003 and by the fellowship SFRH/BPD/70574/2010, SFRH/BPD/76606/2011 funded by FCT and POPH/FSE (EC). PF further acknowledges support from Funda\c{c}\~ao para a Ci\^encia e a Tecnologia (FCT) in the form of an exploratory project of reference IF/01037/2013CP1191/CT0001. L.N. has been funded by the Interuniversity Attraction Poles Programme initiated by the Belgian Science Policy Office through the Planet Topers alliance and by the Deutsche Forschungsgemeinschaft (SFB-TRR 170). C.M. acknowledges the support from the Swiss National Science Foundation under grant BSSGI0$\_$155816 ``PlanetsInTime''. Parts of this work have been carried out within the frame of the National Center for Competence in Research PlanetS supported by the SNSF. This work results within the collaboration of the COST Action TD 1308.

\end{acknowledgements}

%----------------------------------------------------------------------------------------
%       Bibliography
%----------------------------------------------------------------------------------------
\bibliographystyle{aa}
\bibliography{santos_bibliography}

\begin{thebibliography}{93}
\expandafter\ifx\csname natexlab\endcsname\relax\def\natexlab#1{#1}\fi

\bibitem[{{Adibekyan} {et~al.}(2016{\natexlab{a}}){Adibekyan}, {Delgado-Mena},
  {Figueira}, {Sousa}, {Santos}, {Gonz{\'a}lez Hern{\'a}ndez}, {Minchev},
  {Faria}, {Israelian}, {Harutyunyan}, {Su{\'a}rez-Andr{\'e}s}, \&
  {Hakobyan}}]{Adibekyan-2016}
{Adibekyan}, V., {Delgado-Mena}, E., {Figueira}, P., {et~al.}
  2016{\natexlab{a}}, \aap, 592, A87

\bibitem[{{Adibekyan} {et~al.}(2016{\natexlab{b}}){Adibekyan}, {Figueira}, \&
  {Santos}}]{Adibekyan-2016b}
{Adibekyan}, V., {Figueira}, P., \& {Santos}, N.~C. 2016{\natexlab{b}}, Origins
  of Life and Evolution of the Biosphere, 46, 351

\bibitem[{{Adibekyan} {et~al.}(2015){Adibekyan}, {Santos}, {Figueira}, {Dorn},
  {Sousa}, {Delgado-Mena}, {Israelian}, {Hakobyan}, \&
  {Mordasini}}]{Adibekyan-2015b}
{Adibekyan}, V., {Santos}, N.~C., {Figueira}, P., {et~al.} 2015, \aap, 581, L2

\bibitem[{{Adibekyan} {et~al.}(2012{\natexlab{a}}){Adibekyan}, {Delgado Mena},
  {Sousa}, {Santos}, {Israelian}, {Gonz{\'a}lez Hern{\'a}ndez}, {Mayor}, \&
  {Hakobyan}}]{Adibekyan-2012}
{Adibekyan}, V.~Z., {Delgado Mena}, E., {Sousa}, S.~G., {et~al.}
  2012{\natexlab{a}}, A\&A, 547, A36

\bibitem[{{Adibekyan} {et~al.}(2013{\natexlab{a}}){Adibekyan}, {Figueira},
  {Santos}, {Hakobyan}, {Sousa}, {Pace}, {Delgado Mena}, {Robin}, {Israelian},
  \& {Gonz{\'a}lez Hern{\'a}ndez}}]{Adibekyan-2013b}
{Adibekyan}, V.~Z., {Figueira}, P., {Santos}, N.~C., {et~al.}
  2013{\natexlab{a}}, \aap, 554, A44

\bibitem[{{Adibekyan} {et~al.}(2013{\natexlab{b}}){Adibekyan}, {Figueira},
  {Santos}, {Mortier}, {Mordasini}, {Delgado Mena}, {Sousa}, {Correia},
  {Israelian}, \& {Oshagh}}]{Adibekyan-2013}
{Adibekyan}, V.~Z., {Figueira}, P., {Santos}, N.~C., {et~al.}
  2013{\natexlab{b}}, A\&A, 560, A51

\bibitem[{{Adibekyan} {et~al.}(2011{\natexlab{a}}){Adibekyan}, {Santos},
  {Sousa}, \& {Israelian}}]{Adibekyan-2011}
{Adibekyan}, V.~Z., {Santos}, N.~C., {Sousa}, S.~G., \& {Israelian}, G.
  2011{\natexlab{a}}, \aap, 535, L11

\bibitem[{{Adibekyan} {et~al.}(2011{\natexlab{b}}){Adibekyan}, {Santos},
  {Sousa}, \& {Israelian}}]{Adibekyan-11}
{Adibekyan}, V.~Z., {Santos}, N.~C., {Sousa}, S.~G., \& {Israelian}, G.
  2011{\natexlab{b}}, \aap, 535, L11

\bibitem[{{Adibekyan} {et~al.}(2012{\natexlab{b}}){Adibekyan}, {Sousa},
  {Santos}, {Delgado Mena}, {Gonz{\'a}lez Hern{\'a}ndez}, {Israelian}, {Mayor},
  \& {Khachatryan}}]{Adibekyan-2012b}
{Adibekyan}, V.~Z., {Sousa}, S.~G., {Santos}, N.~C., {et~al.}
  2012{\natexlab{b}}, A\&A, 545, A32

\bibitem[{{Asplund} {et~al.}(2009){Asplund}, {Grevesse}, {Sauval}, \&
  {Scott}}]{Asplund-2009}
{Asplund}, M., {Grevesse}, N., {Sauval}, A.~J., \& {Scott}, P. 2009, \araa, 47,
  481

\bibitem[{{Beaug{\'e}} \& {Nesvorn{\'y}}(2013)}]{Beauge-2013}
{Beaug{\'e}}, C. \& {Nesvorn{\'y}}, D. 2013, ApJ, 763, 12

\bibitem[{{Bensby} {et~al.}(2003{\natexlab{a}}){Bensby}, {Feltzing}, \&
  {Lundstr{\" o}m}}]{Bensby-2003}
{Bensby}, T., {Feltzing}, S., \& {Lundstr{\" o}m}, I. 2003{\natexlab{a}}, A\&A,
  410, 527

\bibitem[{{Bensby} {et~al.}(2003{\natexlab{b}}){Bensby}, {Feltzing}, \&
  {Lundstr{\"o}m}}]{Bensby-03}
{Bensby}, T., {Feltzing}, S., \& {Lundstr{\"o}m}, I. 2003{\natexlab{b}}, \aap,
  410, 527

\bibitem[{{Bensby} {et~al.}(2005){Bensby}, {Feltzing}, {Lundstr{\"o}m}, \&
  {Ilyin}}]{Bensby-05}
{Bensby}, T., {Feltzing}, S., {Lundstr{\"o}m}, I., \& {Ilyin}, I. 2005, \aap,
  433, 185

\bibitem[{{Bensby} {et~al.}(2014){Bensby}, {Feltzing}, \& {Oey}}]{Bensby-2014}
{Bensby}, T., {Feltzing}, S., \& {Oey}, M.~S. 2014, \aap, 562, A71

\bibitem[{{Benz} {et~al.}(1988){Benz}, {Slattery}, \& {Cameron}}]{Benz-1988}
{Benz}, W., {Slattery}, W.~L., \& {Cameron}, A.~G.~W. 1988, \icarus, 74, 516

\bibitem[{{Bertran de Lis} {et~al.}(2015){Bertran de Lis}, {Delgado Mena},
  {Adibekyan}, {Santos}, \& {Sousa}}]{BertrandeLis-2015}
{Bertran de Lis}, S., {Delgado Mena}, E., {Adibekyan}, V.~Z., {Santos}, N.~C.,
  \& {Sousa}, S.~G. 2015, \aap, 576, A89

\bibitem[{{Bond} {et~al.}(2010){Bond}, {O'Brien}, \& {Lauretta}}]{Bond-2010}
{Bond}, J.~C., {O'Brien}, D.~P., \& {Lauretta}, D.~S. 2010, ApJ, 715, 1050

\bibitem[{{Bovy} {et~al.}(2012){Bovy}, {Rix}, {Liu}, {Hogg}, {Beers}, \&
  {Lee}}]{Bovy-12}
{Bovy}, J., {Rix}, H.-W., {Liu}, C., {et~al.} 2012, \apj, 753, 148

\bibitem[{{Brewer} \& {Fischer}(2016)}]{Brewer-2016}
{Brewer}, J.~M. \& {Fischer}, D.~A. 2016, \apj, 831, 20

\bibitem[{{Buchhave} {et~al.}(2012){Buchhave}, {Latham}, {Johansen},
  {Bizzarro}, {Torres}, {Rowe}, {Batalha}, {Borucki}, {Brugamyer}, {Caldwell},
  {Bryson}, {Ciardi}, {Cochran}, {Endl}, {Esquerdo}, {Ford}, {Geary},
  {Gilliland}, {Hansen}, {Isaacson}, {Laird}, {Lucas}, {Marcy}, {Morse},
  {Robertson}, {Shporer}, {Stefanik}, {Still}, \& {Quinn}}]{Buchhave-2012}
{Buchhave}, L.~A., {Latham}, D.~W., {Johansen}, A., {et~al.} 2012, Nature, 486,
  375

\bibitem[{{Burgasser} \& {Mamajek}(2017)}]{Burgasser-2017}
{Burgasser}, A.~J. \& {Mamajek}, E.~E. 2017, \apj, 845, 110

\bibitem[{{Buser}(2000)}]{Buser-2000}
{Buser}, R. 2000, Science, 287, 69

\bibitem[{{Carigi} {et~al.}(2013){Carigi}, {Garc{\'{\i}}a-Rojas}, \&
  {Meneses-Goytia}}]{Carigi-2013}
{Carigi}, L., {Garc{\'{\i}}a-Rojas}, J., \& {Meneses-Goytia}, S. 2013, \rmxaa,
  49, 253

\bibitem[{{Carter-Bond} {et~al.}(2012){Carter-Bond}, {O'Brien}, \&
  {Raymond}}]{Carter-Bond-2012}
{Carter-Bond}, J.~C., {O'Brien}, D.~P., \& {Raymond}, S.~N. 2012, \apj, 760, 44

\bibitem[{{Dawson} \& {Murray-Clay}(2013)}]{Dawson-2013}
{Dawson}, R.~I. \& {Murray-Clay}, R.~A. 2013, ApJ, 767, L24

\bibitem[{{Delgado Mena} {et~al.}(2010){Delgado Mena}, {Israelian},
  {Gonz{\'a}lez Hern{\'a}ndez}, {Bond}, {Santos}, {Udry}, \&
  {Mayor}}]{Delgado-Mena-2010}
{Delgado Mena}, E., {Israelian}, G., {Gonz{\'a}lez Hern{\'a}ndez}, J.~I.,
  {et~al.} 2010, \apj, 725, 2349

\bibitem[{{Dittmann} {et~al.}(2017){Dittmann}, {Irwin}, {Charbonneau},
  {Bonfils}, {Astudillo-Defru}, {Haywood}, {Berta-Thompson}, {Newton},
  {Rodriguez}, {Winters}, {Tan}, {Almenara}, {Bouchy}, {Delfosse}, {Forveille},
  {Lovis}, {Murgas}, {Pepe}, {Santos}, {Udry}, {W{\"u}nsche}, {Esquerdo},
  {Latham}, \& {Dressing}}]{Dittmann-2017}
{Dittmann}, J.~A., {Irwin}, J.~M., {Charbonneau}, D., {et~al.} 2017, \nat, 544,
  333

\bibitem[{{Dorn} {et~al.}(2017){Dorn}, {Hinkel}, \& {Venturini}}]{Dorn-2017}
{Dorn}, C., {Hinkel}, N.~R., \& {Venturini}, J. 2017, \aap, 597, A38

\bibitem[{{Dorn} {et~al.}(2015){Dorn}, {Khan}, {Heng}, {Connolly}, {Alibert},
  {Benz}, \& {Tackley}}]{Dorn-2015}
{Dorn}, C., {Khan}, A., {Heng}, K., {et~al.} 2015, \aap, 577, A83

\bibitem[{{Drake} \& {Righter}(2002)}]{Drake-2002}
{Drake}, M.~J. \& {Righter}, K. 2002, \nat, 416, 39

\bibitem[{{Fischer} \& {Valenti}(2005)}]{Fischer-2005}
{Fischer}, D.~A. \& {Valenti}, J. 2005, ApJ, 622, 1102

\bibitem[{Fischer {et~al.}(2017)Fischer, Campbell, \& Ciesla}]{Fischer-2017}
Fischer, R.~A., Campbell, A.~J., \& Ciesla, F.~J. 2017, Earth and Planetary
  Science Letters, 458, 252

\bibitem[{{Fortney}(2012)}]{Fortney-2012}
{Fortney}, J.~J. 2012, \apjl, 747, L27

\bibitem[{{Fortney} {et~al.}(2007){Fortney}, {Marley}, \&
  {Barnes}}]{Fortney-2007}
{Fortney}, J.~J., {Marley}, M.~S., \& {Barnes}, J.~W. 2007, ApJ, 659, 1661

\bibitem[{{Frank} {et~al.}(2014){Frank}, {Meyer}, \& {Mojzsis}}]{Frank-2014}
{Frank}, E.~A., {Meyer}, B.~S., \& {Mojzsis}, S.~J. 2014, \icarus, 243, 274

\bibitem[{{Fridlund} {et~al.}(2017){Fridlund}, {Gaidos}, {Barrag{\'a}n},
  {Persson}, {Gandolfi}, {Cabrera}, {Hirano}, {Kuzuhara}, {Csizmadia}, {Nowak},
  {Endl}, {Grziwa}, {Korth}, {Pfaff}, {Bitsch}, {Johansen}, {Mustill},
  {Davies}, {Deeg}, {Palle}, {Cochran}, {Eigm{\"u}ller}, {Erikson}, {Guenther},
  {Hatzes}, {Kiilerich}, {Kudo}, {MacQueen}, {Narita}, {Nespral},
  {P{\"a}tzold}, {Prieto-Arranz}, {Rauer}, \& {Van Eylen}}]{Fridlund-2017}
{Fridlund}, M., {Gaidos}, E., {Barrag{\'a}n}, O., {et~al.} 2017, ArXiv e-prints

\bibitem[{{Fuhrmann}(2008)}]{Fuhrmann-08}
{Fuhrmann}, K. 2008, \mnras, 384, 173

\bibitem[{{Gillon} {et~al.}(2017){Gillon}, {Triaud}, {Demory}, {Jehin}, {Agol},
  {Deck}, {Lederer}, {de Wit}, {Burdanov}, {Ingalls}, {Bolmont}, {Leconte},
  {Raymond}, {Selsis}, {Turbet}, {Barkaoui}, {Burgasser}, {Burleigh}, {Carey},
  {Chaushev}, {Copperwheat}, {Delrez}, {Fernandes}, {Holdsworth}, {Kotze}, {Van
  Grootel}, {Almleaky}, {Benkhaldoun}, {Magain}, \& {Queloz}}]{Gillon-2017}
{Gillon}, M., {Triaud}, A.~H.~M.~J., {Demory}, B.-O., {et~al.} 2017, \nat, 542,
  456

\bibitem[{{Gilmore} \& {Reid}(1983)}]{Gilmore-1983}
{Gilmore}, G. \& {Reid}, N. 1983, \mnras, 202, 1025

\bibitem[{{Gonzalez}(2014)}]{Gonzalez-2014}
{Gonzalez}, G. 2014, \mnras, 443, 393

\bibitem[{{Guenther} {et~al.}(2017){Guenther}, {Barragan}, {Dai}, {Gandolfi},
  {Hirano}, {Fridlund}, {Fossati}, {Korth}, {Arraz-Prieto}, {Nespral},
  {Antoniciello}, {Deeg}, {Hjorth}, {Grziwa}, {Albrecht}, {Hatzes}, {Rauer},
  {Csizmadia}, {Smith}, {Cabrera}, {Narita}, {Arriagada}, {Burt}, {Butler},
  {Cochran}, {Crane}, {Eigmueller}, {Erikson}, {Johnson}, {Kiilerich},
  {Kubyshkina}, {Kunz}, {Palle}, {Persson}, {Paetzold}, {Prieto-Arranz},
  {Sato}, {Shectman}, {Teske}, {Thompson}, {Van Eylen}, {Nowak}, {Vanderburg},
  {Winn}, \& {Wittenmyer}}]{Guenther-2017}
{Guenther}, E.~W., {Barragan}, O., {Dai}, F., {et~al.} 2017, ArXiv e-prints

\bibitem[{{Guillot} {et~al.}(2006){Guillot}, {Santos}, {Pont}, {Iro}, {Melo},
  \& {Ribas}}]{Guillot-2006}
{Guillot}, T., {Santos}, N.~C., {Pont}, F., {et~al.} 2006, A\&A, 453, L21

\bibitem[{{Hayden} {et~al.}(2015){Hayden}, {Bovy}, {Holtzman}, {Nidever},
  {Bird}, {Weinberg}, {Andrews}, {Majewski}, {Allende Prieto}, {Anders},
  {Beers}, {Bizyaev}, {Chiappini}, {Cunha}, {Frinchaboy},
  {Garc{\'{\i}}a-Her{\'n}andez}, {Garc{\'{\i}}a P{\'e}rez}, {Girardi},
  {Harding}, {Hearty}, {Johnson}, {M{\'e}sz{\'a}ros}, {Minchev}, {O'Connell},
  {Pan}, {Robin}, {Schiavon}, {Schneider}, {Schultheis}, {Shetrone},
  {Skrutskie}, {Steinmetz}, {Smith}, {Wilson}, {Zamora}, \&
  {Zasowski}}]{Hayden-2015}
{Hayden}, M.~R., {Bovy}, J., {Holtzman}, J.~A., {et~al.} 2015, \apj, 808, 132

\bibitem[{{Haywood}(2008)}]{Haywood-2008}
{Haywood}, M. 2008, A\&A, 482, 673

\bibitem[{{Haywood} {et~al.}(2013{\natexlab{a}}){Haywood}, {Di Matteo},
  {Lehnert}, {Katz}, \& {G{\'o}mez}}]{Haywood-2013}
{Haywood}, M., {Di Matteo}, P., {Lehnert}, M.~D., {Katz}, D., \& {G{\'o}mez},
  A. 2013{\natexlab{a}}, \aap, 560, A109

\bibitem[{{Haywood} {et~al.}(2013{\natexlab{b}}){Haywood}, {Di Matteo},
  {Lehnert}, {Katz}, \& {G{\'o}mez}}]{Haywood-13}
{Haywood}, M., {Di Matteo}, P., {Lehnert}, M.~D., {Katz}, D., \& {G{\'o}mez},
  A. 2013{\natexlab{b}}, \aap, 560, A109

\bibitem[{{Haywood} {et~al.}(2016){Haywood}, {Lehnert}, {Di Matteo}, {Snaith},
  {Schultheis}, {Katz}, \& {G{\'o}mez}}]{Haywood-2016}
{Haywood}, M., {Lehnert}, M.~D., {Di Matteo}, P., {et~al.} 2016, \aap, 589, A66

\bibitem[{{Israelian} {et~al.}(2001){Israelian}, {Santos}, {Mayor}, \&
  {Rebolo}}]{Israelian-2001}
{Israelian}, G., {Santos}, N.~C., {Mayor}, M., \& {Rebolo}, R. 2001, Nature,
  411, 163

\bibitem[{{Ivezi{\'c}} {et~al.}(2012){Ivezi{\'c}}, {Beers}, \&
  {Juri{\'c}}}]{Ivezic-12}
{Ivezi{\'c}}, {\v Z}., {Beers}, T.~C., \& {Juri{\'c}}, M. 2012, \araa, 50, 251

\bibitem[{{Johnson} {et~al.}(2007){Johnson}, {Butler}, {Marcy}, {Fischer},
  {Vogt}, {Wright}, \& {Peek}}]{Johnson-2007}
{Johnson}, J.~A., {Butler}, R.~P., {Marcy}, G.~W., {et~al.} 2007, ApJ, 670, 833

\bibitem[{{Juri{\'c}} {et~al.}(2008){Juri{\'c}}, {Ivezi{\'c}}, {Brooks},
  {Lupton}, {Schlegel}, {Finkbeiner}, {Padmanabhan}, {Bond}, {Sesar},
  {Rockosi}, {Knapp}, {Gunn}, {Sumi}, {Schneider}, {Barentine}, {Brewington},
  {Brinkmann}, {Fukugita}, {Harvanek}, {Kleinman}, {Krzesinski}, {Long},
  {Neilsen}, {Nitta}, {Snedden}, \& {York}}]{Juric-08}
{Juri{\'c}}, M., {Ivezi{\'c}}, {\v Z}., {Brooks}, A., {et~al.} 2008, \apj, 673,
  864

\bibitem[{Khan \& Connolly(2008)}]{Khan-2008}
Khan, A. \& Connolly, J. A.~D. 2008, Journal of Geophysical Research: Planets,
  113, n/a, e07003

\bibitem[{{Kordopatis} {et~al.}(2015){Kordopatis}, {Wyse}, {Gilmore},
  {Recio-Blanco}, {de Laverny}, {Hill}, {Adibekyan}, {Heiter}, {Minchev},
  {Famaey}, {Bensby}, {Feltzing}, {Guiglion}, {Korn}, {Mikolaitis},
  {Schultheis}, {Vallenari}, {Bayo}, {Carraro}, {Flaccomio}, {Franciosini},
  {Hourihane}, {Jofr{\'e}}, {Koposov}, {Lardo}, {Lewis}, {Lind}, {Magrini},
  {Morbidelli}, {Pancino}, {Randich}, {Sacco}, {Worley}, \&
  {Zaggia}}]{Kordopatis-15}
{Kordopatis}, G., {Wyse}, R.~F.~G., {Gilmore}, G., {et~al.} 2015, \aap, 582,
  A122

\bibitem[{{Korenaga}(2010)}]{Korenaga-2010}
{Korenaga}, J. 2010, \apjl, 725, L43

\bibitem[{{L{\'e}ger} {et~al.}(2004){L{\'e}ger}, {Selsis}, {Sotin}, {Guillot},
  {Despois}, {Mawet}, {Ollivier}, {Lab{\`e}que}, {Valette}, {Brachet},
  {Chazelas}, \& {Lammer}}]{Leger-2004}
{L{\'e}ger}, A., {Selsis}, F., {Sotin}, C., {et~al.} 2004, \icarus, 169, 499

\bibitem[{{Lineweaver} {et~al.}(2004){Lineweaver}, {Fenner}, \&
  {Gibson}}]{Lineweaver-2004}
{Lineweaver}, C.~H., {Fenner}, Y., \& {Gibson}, B.~K. 2004, Science, 303, 59

\bibitem[{{Lodders}(2003)}]{Lodders-2003}
{Lodders}, K. 2003, ApJ, 591, 1220

\bibitem[{{Lodders} \& {Fegley}(1998)}]{Lodders-1998}
{Lodders}, K. \& {Fegley}, B. 1998, {The planetary scientist's companion /
  Katharina Lodders, Bruce Fegley.}

\bibitem[{{Luger} {et~al.}(2015){Luger}, {Barnes}, {Lopez}, {Fortney},
  {Jackson}, \& {Meadows}}]{Luger-2015}
{Luger}, R., {Barnes}, R., {Lopez}, E., {et~al.} 2015, Astrobiology, 15, 57

\bibitem[{{Marcus} {et~al.}(2010{\natexlab{a}}){Marcus}, {Sasselov},
  {Hernquist}, \& {Stewart}}]{Marcus-2010a}
{Marcus}, R.~A., {Sasselov}, D., {Hernquist}, L., \& {Stewart}, S.~T.
  2010{\natexlab{a}}, \apjl, 712, L73

\bibitem[{{Marcus} {et~al.}(2010{\natexlab{b}}){Marcus}, {Sasselov}, {Stewart},
  \& {Hernquist}}]{Marcus-2010b}
{Marcus}, R.~A., {Sasselov}, D., {Stewart}, S.~T., \& {Hernquist}, L.
  2010{\natexlab{b}}, \apjl, 719, L45

\bibitem[{{Mayor} {et~al.}(2014){Mayor}, {Lovis}, \& {Santos}}]{Mayor-2014}
{Mayor}, M., {Lovis}, C., \& {Santos}, N.~C. 2014, \nat, 513, 328

\bibitem[{{McDounough} \& {Sun}(1995)}]{McDounough-1995}
{McDounough}, W.~F. \& {Sun}, S.-s. 1995, Chemical Geology, 120, 223

\bibitem[{{Mordasini} {et~al.}(2012{\natexlab{a}}){Mordasini}, {Alibert},
  {Benz}, {Klahr}, \& {Henning}}]{Mordasini-2012}
{Mordasini}, C., {Alibert}, Y., {Benz}, W., {Klahr}, H., \& {Henning}, T.
  2012{\natexlab{a}}, A\&A, 541, A97

\bibitem[{{Mordasini} {et~al.}(2012{\natexlab{b}}){Mordasini}, {Alibert},
  {Georgy}, {Dittkrist}, {Klahr}, \& {Henning}}]{Mordasini-2012b}
{Mordasini}, C., {Alibert}, Y., {Georgy}, C., {et~al.} 2012{\natexlab{b}},
  \aap, 547, A112

\bibitem[{{Morgan} \& {Anders}(1980)}]{Morgan-1980}
{Morgan}, J.~W. \& {Anders}, E. 1980, Proceedings of the National Academy of
  Science, 77, 6973

\bibitem[{{Motalebi} {et~al.}(2015){Motalebi}, {Udry}, {Gillon}, {Lovis},
  {S{\'e}gransan}, {Buchhave}, {Demory}, {Malavolta}, {Dressing}, {Sasselov},
  {Rice}, {Charbonneau}, {Collier Cameron}, {Latham}, {Molinari}, {Pepe},
  {Affer}, {Bonomo}, {Cosentino}, {Dumusque}, {Figueira}, {Fiorenzano},
  {Gettel}, {Harutyunyan}, {Haywood}, {Johnson}, {Lopez}, {Lopez-Morales},
  {Mayor}, {Micela}, {Mortier}, {Nascimbeni}, {Philips}, {Piotto}, {Pollacco},
  {Queloz}, {Sozzetti}, {Vanderburg}, \& {Watson}}]{Motalebi-2015}
{Motalebi}, F., {Udry}, S., {Gillon}, M., {et~al.} 2015, \aap, 584, A72

\bibitem[{{Navarro} {et~al.}(2011){Navarro}, {Abadi}, {Venn}, {Freeman}, \&
  {Anguiano}}]{Navarro-11}
{Navarro}, J.~F., {Abadi}, M.~G., {Venn}, K.~A., {Freeman}, K.~C., \&
  {Anguiano}, B. 2011, \mnras, 412, 1203

\bibitem[{{Nissen} \& {Schuster}(2010)}]{Nissen-2010}
{Nissen}, P.~E. \& {Schuster}, W.~J. 2010, \aap, 511, L10

\bibitem[{{Noack} {et~al.}(2014){Noack}, {Godolt}, {von Paris}, {Plesa},
  {Stracke}, {Breuer}, \& {Rauer}}]{Noack-2014}
{Noack}, L., {Godolt}, M., {von Paris}, P., {et~al.} 2014, \planss, 98, 14

\bibitem[{Noack {et~al.}(2017)Noack, Rivoldini, \& Hoolst}]{Noack-2017}
Noack, L., Rivoldini, A., \& Hoolst, T.~V. 2017, Physics of the Earth and
  Planetary Interiors, 269, 40

\bibitem[{{{\"O}nehag} {et~al.}(2014){{\"O}nehag}, {Gustafsson}, \&
  {Korn}}]{Onehag-2014}
{{\"O}nehag}, A., {Gustafsson}, B., \& {Korn}, A. 2014, \aap, 562, A102

\bibitem[{{Prantzos}(2008)}]{Prantzos-2008}
{Prantzos}, N. 2008, \ssr, 135, 313

\bibitem[{{Recio-Blanco} {et~al.}(2014){Recio-Blanco}, {de Laverny},
  {Kordopatis}, {Helmi}, {Hill}, {Gilmore}, {Wyse}, {Adibekyan}, {Randich},
  {Asplund}, {Feltzing}, {Jeffries}, {Micela}, {Vallenari}, {Alfaro}, {Allende
  Prieto}, {Bensby}, {Bragaglia}, {Flaccomio}, {Koposov}, {Korn}, {Lanzafame},
  {Pancino}, {Smiljanic}, {Jackson}, {Lewis}, {Magrini}, {Morbidelli},
  {Prisinzano}, {Sacco}, {Worley}, {Hourihane}, {Bergemann}, {Costado},
  {Heiter}, {Joffre}, {Lardo}, {Lind}, \& {Maiorca}}]{Recio-Blanco-14}
{Recio-Blanco}, A., {de Laverny}, P., {Kordopatis}, G., {et~al.} 2014, \aap,
  567, A5

\bibitem[{Rubie {et~al.}(2015)Rubie, Jacobson, Morbidelli, O’Brien, Young,
  de~Vries, Nimmo, Palme, \& Frost}]{Rubie-2015}
Rubie, D., Jacobson, S., Morbidelli, A., {et~al.} 2015, Icarus, 248, 89

\bibitem[{{Sanloup} {et~al.}(1999){Sanloup}, {Jambon}, \&
  {Gillet}}]{Sanloup-1999}
{Sanloup}, C., {Jambon}, A., \& {Gillet}, P. 1999, Physics of the Earth and
  Planetary Interiors, 112, 43

\bibitem[{{Santos} {et~al.}(2015){Santos}, {Adibekyan}, {Mordasini}, {Benz},
  {Delgado-Mena}, {Dorn}, {Buchhave}, {Figueira}, {Mortier}, {Pepe},
  {Santerne}, {Sousa}, \& {Udry}}]{Santos-2015a}
{Santos}, N.~C., {Adibekyan}, V., {Mordasini}, C., {et~al.} 2015, \aap, 580,
  L13

\bibitem[{{Santos} {et~al.}(2004){Santos}, {Israelian}, \&
  {Mayor}}]{Santos-2004b}
{Santos}, N.~C., {Israelian}, G., \& {Mayor}, M. 2004, A\&A, 415, 1153

\bibitem[{{Santos} {et~al.}(2003){Santos}, {Israelian}, {Mayor}, {Rebolo}, \&
  {Udry}}]{Santos-2003}
{Santos}, N.~C., {Israelian}, G., {Mayor}, M., {Rebolo}, R., \& {Udry}, S.
  2003, A\&A, 398, 363

\bibitem[{{Seager} {et~al.}(2007){Seager}, {Kuchner}, {Hier-Majumder}, \&
  {Militzer}}]{Seager-2007}
{Seager}, S., {Kuchner}, M., {Hier-Majumder}, C.~A., \& {Militzer}, B. 2007,
  \apj, 669, 1279

\bibitem[{{Simpson}(2017)}]{Simpson-2017}
{Simpson}, F. 2017, \mnras, 468, 2803

\bibitem[{{Snaith} {et~al.}(2015){Snaith}, {Haywood}, {Di Matteo}, {Lehnert},
  {Combes}, {Katz}, \& {G{\'o}mez}}]{Snaith-2015}
{Snaith}, O., {Haywood}, M., {Di Matteo}, P., {et~al.} 2015, \aap, 578, A87

\bibitem[{{Sotin} {et~al.}(2007){Sotin}, {Grasset}, \& {Mocquet}}]{Sotin-2007}
{Sotin}, C., {Grasset}, O., \& {Mocquet}, A. 2007, \icarus, 191, 337

\bibitem[{{Sousa} {et~al.}(2011{\natexlab{a}}){Sousa}, {Santos}, {Israelian},
  {Lovis}, {Mayor}, {Silva}, \& {Udry}}]{Sousa-2011b}
{Sousa}, S.~G., {Santos}, N.~C., {Israelian}, G., {et~al.} 2011{\natexlab{a}},
  A\&A, 526, A99

\bibitem[{{Sousa} {et~al.}(2011{\natexlab{b}}){Sousa}, {Santos}, {Israelian},
  {Mayor}, \& {Udry}}]{Sousa-2011}
{Sousa}, S.~G., {Santos}, N.~C., {Israelian}, G., {Mayor}, M., \& {Udry}, S.
  2011{\natexlab{b}}, A\&A, 533, A141+

\bibitem[{{Sousa} {et~al.}(2008){Sousa}, {Santos}, {Mayor}, {Udry},
  {Casagrande}, {Israelian}, {Pepe}, {Queloz}, \& {Monteiro}}]{Sousa-2008}
{Sousa}, S.~G., {Santos}, N.~C., {Mayor}, M., {et~al.} 2008, A\&A, 487, 373

\bibitem[{{Spitoni} {et~al.}(2017){Spitoni}, {Gioannini}, \&
  {Matteucci}}]{Spitoni-2017}
{Spitoni}, E., {Gioannini}, L., \& {Matteucci}, F. 2017, ArXiv e-prints

\bibitem[{{Su{\'a}rez-Andr{\'e}s} {et~al.}(2017){Su{\'a}rez-Andr{\'e}s},
  {Israelian}, {Gonz{\'a}lez Hern{\'a}ndez}, {Adibekyan}, {Delgado Mena},
  {Santos}, \& {Sousa}}]{Suarez-Andres-2017}
{Su{\'a}rez-Andr{\'e}s}, L., {Israelian}, G., {Gonz{\'a}lez Hern{\'a}ndez},
  J.~I., {et~al.} 2017, \aap, 599, A96

\bibitem[{{Thiabaud} {et~al.}(2015){Thiabaud}, {Marboeuf}, {Alibert}, {Leya},
  \& {Mezger}}]{Thiabaud-2015}
{Thiabaud}, A., {Marboeuf}, U., {Alibert}, Y., {Leya}, I., \& {Mezger}, K.
  2015, \aap, 580, A30

\bibitem[{{Unterborn} \& {Panero}(2016)}]{Unterborn-2016}
{Unterborn}, C.~T. \& {Panero}, W.~R. 2016, ArXiv e-prints

\bibitem[{{Wang} \& {Fischer}(2015)}]{Wang-2015}
{Wang}, J. \& {Fischer}, D.~A. 2015, \aj, 149, 14

\bibitem[{{Zhu} {et~al.}(2016){Zhu}, {Wang}, \& {Huang}}]{Zhu-2016}
{Zhu}, W., {Wang}, J., \& {Huang}, C. 2016, \apj, 832, 196

\end{thebibliography}

%----------------------------------------------------------------------------------
%       Appendices
%----------------------------------------------------------------------------------
%\appendix

%\section{Flux received by the ring}  \label{app:flux}

% Here, we detail the computation of the flux $F_r(\phi)$ received by the

\begin{appendix}
\label{appendix1}

\section{Notes on Galactic populations} 
\label{appendix1A}

The division between the disc and halo was identified long ago, but the thick disc was discovered far more recently by \citet[][]{Gilmore-1983} by analysing the stellar density distribution as a function of distance from the Galactic plane. The thin- and thick-disc populations have different kinematics and chemical properties. 
{Generally, the thick disc is thought to be composed of relatively old stars \citep[e.g.][]{Bensby-05,Fuhrmann-08,Adibekyan-11}
that usually are metal poor and $\alpha$-element enhanced \citep[e.g.][]{Adibekyan-2013b,Recio-Blanco-14}, and for which the stellar number density has a large scale-height and short scale-length\footnote{We note, however, that previous studies have obtained longer scale-lengths for the stellar density profile of the thick disc \citep[e.g.][]{Juric-08}.} \citep[e.g.][]{Bovy-12}.}
%Generally, the thick disk is thought to be composed of relatively old {\bf stars \citep[e.g.][]{Bensby-05,Fuhrmann-08,Adibekyan-11}, usually metal-poor, and alpha-enhanced \citep[e.g.][]{Adibekyan-2013b,Recio-Blanco-14}, and moving in Galactic orbits with a large scale height and long-scale length \citep[e.g.][]{Juric-08, Bovy-12}.} 
Most stars {in the solar neighbourhood}\footnote{This value is around $\sim$50\% if we consider the whole Galaxy \citep[e.g.][]{Snaith-2015}.} are members of the younger thin-disc component,  and they range in [Fe/H] from $\sim$-0.8 up to $\sim$+0.5 dex \citep[][]{Kordopatis-15,Adibekyan-2013b}. Thick-disc and halo stars typically have lower metallicities than their thin-disc counterparts. 

There is no obvious predetermined way to distinguish between different stellar populations in the solar neighbourhood. However, since chemistry is a relatively more stable property of a star than its spatial positions and kinematics, it is becoming increasingly clear that a dissection of the Galactic discs based only on stellar abundances is superior to kinematic separation \citep[see][]{Navarro-11,Adibekyan-11}. Stellar ages can also be effectively used to separate the thin- and thick-disc stars  \citep[e.g.][]{Haywood-13}, although they are very difficult to obtain in high precision. Halo stars are
commonly identified using dynamical approaches \citep[e.g.][]{Bensby-03} since these stars share similar chemical properties with their thick-disc counterparts. The kinematical and chemical properties (in particular the abundance ratios) of these three populations reflect their origin, age, and the galactic formation process \citep[e.g.][]{Ivezic-12}.

{We refer to Fig.\,1 of \citet[][]{Buser-2000} for a good scheme of the different Galactic populations.}

\section{Our model equations}
\label{appendix1B}

{
Our model is based on the following stoichiometric relations.
When N$_{Mg}>$N$_{Si}$ \citep[see e.g.][]{Bond-2010,Thiabaud-2015,Unterborn-2016},
\vspace{-0.15truecm}
\begin{eqnarray}
N_{\rm O}&=&N_{\rm H_2O}+3 N_{\rm MgSiO_3}+4 N_{\rm Mg_2SiO_4}\\[-0.1truecm]
N_{\rm Mg}&=&N_{\rm MgSiO_3}+2  N_{\rm Mg_2SiO_4}\\[-0.1truecm]
N_{\rm Si}&=&N_{\rm MgSiO_3}+  N_{\rm Mg_2SiO_4}\\[-0.1truecm]
N_{\rm C} &=& N_{\rm CH_4}
,\end{eqnarray}
otherwise, when N$_{Mg}\leq$N$_{Si}$,
\begin{eqnarray}
N_{\rm O}&=&N_{\rm H_2O}+3 N_{\rm MgSiO_3}+2 N_{\rm SiO_2}\\[-0.1truecm]
N_{\rm Mg}&=&N_{\rm MgSiO_3}\\[-0.1truecm]
N_{\rm Si}&=&N_{\rm MgSiO_3}+  N_{\rm SiO_2}\\[-0.1truecm]
N_{\rm C} &=& N_{\rm CH_4}
.\end{eqnarray}
Inverting these equations and adding the observed stellar abundances allows us to derive the ratios analysed in this work.
}

\section{Predicting the iron mass fraction} 
\label{appendix1C}

\begin{figure}
\begin{center}
\begin{tabular}{c}
\includegraphics[angle=0,width=1\linewidth]{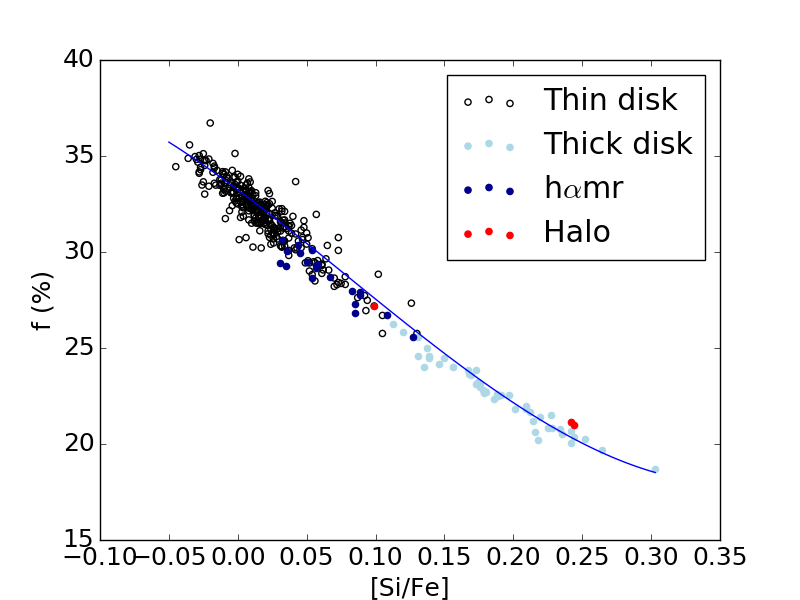}\\[-0.0truecm]
\includegraphics[angle=0,width=1\linewidth]{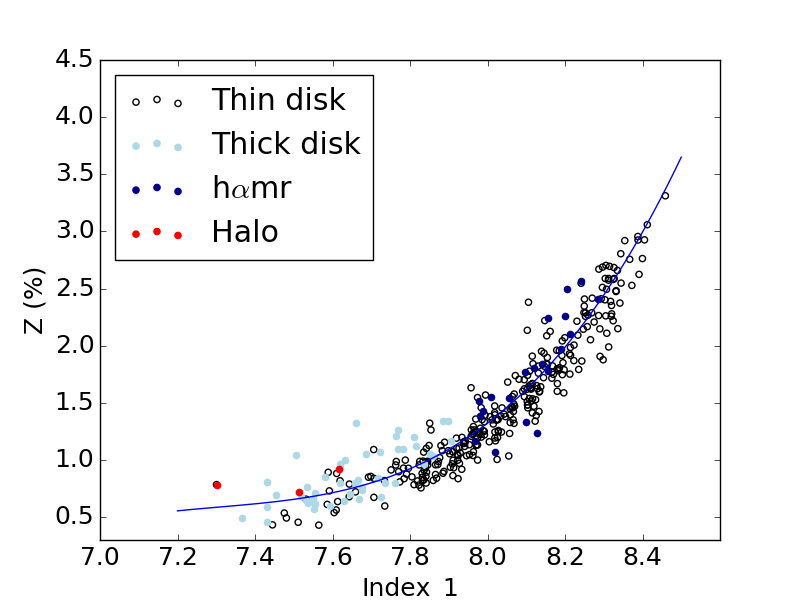}\\[-0.0truecm]
\includegraphics[angle=0,width=1\linewidth]{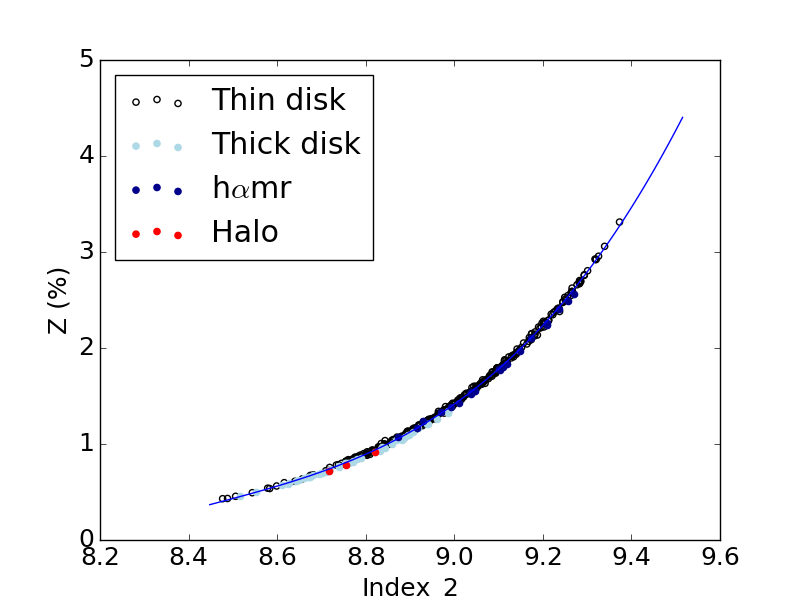}\\[-0.0truecm]
\end{tabular}
\end{center}
\vspace{-0.5cm}
\caption{Upper panel: values of f$_{iron}$ (upper panel) as a function of [Si/Fe]. Middle and lower panels: $Z$  as a function of two different index values as defined in the text. Different galactic populations are denoted with different symbols. The lines denote linear cubic fits to the data.}
\label{fig:correl}
\end{figure}

In Fig.\,\ref{fig:correl} (upper panel) we present the derived $f_{\rm iron}$ for all the stars in our sample as a function of [Si/Fe]. We divided the stars into different populations, following the definition discussed in Sect.\,\ref{sec:data}. The plot illustrates the dependence of our model on the different chemical abundances. In particular, it shows that $f_{\rm iron}$ is strongly dependent on the [Si/Fe] ratio\footnote{The same correlation is also obtained using the [Mg/Fe] ratio, as these two are tightly related. However, since there are more Si lines in the spectrum of a solar-type star, precise Si abundances are easier to obtain, and so we favour using Si here.}. A cubic fit to the data provides the relation
\begin{equation}
\begin{aligned}
f = 253.522\,[Si/Fe]^3-61.149\,[Si/Fe]^2\\
 -53.342\,[Si/Fe]+33.240
\end{aligned}
.\end{equation}

This equation can be used to predict the iron mass contents for the planetary building blocks when the values of [Si/Fe] are known.

Similarly, in Fig.\,\ref{fig:correl} (middle panel), we plot the relation between the values of $Z$ and an index defined as

\begin{equation}
\begin{aligned}
Index_1 = \log{ 10^{\epsilon(Mg)}+10^{\epsilon(Si)}+10^{\epsilon(Fe)}  }
\end{aligned}
.\end{equation}

Here $\epsilon$(X) denotes the abundance of a given element. The choice of Mg, Si, and Fe to build this index comes from the fact that these are the easiest to measure in a solar-type star
of the elements we studied here. 

\begin{equation}
\begin{aligned}
Z = 1.596\,Index_1^3-34.971\,Index_1^2\\
 +255.799\,Index_1-623.802
\end{aligned}
.\end{equation}

This relation can be used to estimate the value of $Z$ in any solar neighbourhood star when the abundances of these three elements are known.

Finally, when the abundances for C and O are also available, we can define the index above as

\begin{equation}
\begin{aligned}
Index_2 = \log{ 10^{\epsilon(Mg)}+10^{\epsilon(Si)}+10^{\epsilon(Fe)}+10^{\epsilon(C)}+10^{\epsilon(O)}  }
\end{aligned}
.\end{equation}

In this case, a cubic fit to the data (see Fig.\,\ref{fig:correl}, lower panel) provides a much tighter relation:

\begin{equation}
\begin{aligned}
Z = 2.280\,Index_2^3-57.831\,Index_2^2\\
 +490.140\,Index_2-1387.773
\end{aligned}
.\end{equation}

\end{appendix}

\end{document}